# Structurally Tunable Nonlinear Terahertz Metamaterials using Broadside Coupled Split Ring Resonators

George R. Keiser[1, a)], Nicholas Karl[2], S. Rubaiat Ul Haque[3], Igal Brener[2], Daniel M. Mittleman[4], and Richard D. Averitt[3]

[1]*Department of Physics, Washington College, Chestertown, MD, USA*

[2]*Center for Integrated Nanotechnologies, Sandia National Lab. Albuquerque, NM, USA*

[3]*Department of Physics, University of California at San Diego, La Jolla, CA, USA*

[4]*School of Engineering, Brown University, Providence, RI, USA*

We present an experimental and numerical study of a terahertz metamaterial with a nonlinear response that is controllable via the relative structural positioning of two stacked split ring resonator arrays. The first array is fabricated on an n-doped GaAs substrate and the second array is fabricated vertically above the first using a polyimide spacer layer. Due to GaAs carrier dynamics, the on-resonance terahertz transmission at 0.4 THz varies in a nonlinear manner with incident terahertz power. The second resonator layer dampens this nonlinear response. In samples where the two layers are aligned, the resonance disappears and total nonlinear modulation of the on-resonance transmission decreases. The nonlinear modulation is restored in samples where an alignment offset is imposed between the two resonator arrays. Structurally tunable MMs can therefore act as a design template for tunable nonlinear THz devices by controlling the coupling of confined electric fields to nonlinear phenomena in a complex material substrate or inclusion.

Over the past decade the field of terahertz (THz) science has grown to include the study and engineering of nonlinear optical materials.[1, 2] This growth in nonlinear THz science has been driven by improvements in the ability to generate high power THz pulses in the laboratory. Standard THz generation via photo-conductive antennas or optical rectification (e.g. in ZnTe) produce THz pulses limited to femto Joules of energy.[3] By contrast, the newer techniques of tilted pulse front THz generation (TPFG) in $LiNbO_3$ and 4-wave mixing in air generate pulse energies orders of magnitude higher.[4, 5] In particular, TPFG methods are known to reliably produce THz pulses with micro Joules of energy, corresponding to peak electric field strengths on the order of 300-500 kV/cm, and in some instances up to 1 MV/cm.[6]

The use of high peak power THz pulses makes time resolved nonlinear experiments possible at THz frequencies. Researchers are able to study the THz nonlinear properties of materials on ultrafast timescales

---

[a)] Author to whom correspondence should be addressed. gkeiser2@washcoll.edu

[7] and design novel nonlinear optical elements for THz radiation. [8, 9] Combining THz excitation with X-ray probes allows for time-resolved measurements of nonlinear structural changes in complex materials. [10]

Within the field of nonlinear THz science, plasmonic metamaterials and metasurfaces (MMs) play a critical role. MMs are engineered composites with optical properties that are determined by the geometry and layout of the component sub-wavelength inclusions.[11] MMs have been applied across the electromagnetic spectrum and may be used to precisely control light through manipulation of phase, intensity, and polarization. [12]. This unprecedented level of control over light has led to notable MM demonstrations of left-handed materials [13], electromagnetic cloaking [14], and perfect absorption[15, 16]. Additionally, MMs are interesting from a materials science standpoint due to their enhancement of light-matter interaction [2, 7]. This interaction can be seen in a magnified nonlinear MM response, making MMs an ideal platform to study THz nonlinear behavior and design nonlinear devices.

Nonlinear MM devices have been created at microwave and terahertz frequencies through a variety of methods including incorporating nonlinear lumped circuit elements into the design [17-19], and by using the inherent nonlinear response of the subwavelength inclusions that make up the MM[20]. Another approach used in nonlinear MMs devices makes use of resonant inclusions, such as the split ring resonator (SRR), which confine electric fields to localized regions in the unit cell. Resonant field confinement (FC) enhances the local peak electric field intensity, and may excite a nonlinear response in the MM inclusions or substrate.[7, 21] Many demonstrations of nonlinear MMs based on FC exist, including devices that couple confined fields to charge carrier dynamics in $VO_2$, GaAs, and InAs [21, 22], and devices that couple fields to superconducting phase transitions in both low and high $T_c$ superconductors. [23-25]

Dynamic control over the MM response dramatically extends device capability; and integration of control and tuning is an active area of study for linear and nonlinear MM devices. Two of the most commonly used methods for electrical control of MM properties are via modulation of substrate conductivity [26] and structural design or actuation, for instance via microelectrical mechanical (MEMS) based devices [27, 28]. We recently have shown that the nonlinear response of a THz MM can be modulated through control of substrate conductivity[9]. Yet to date there has been no report on how the nonlinear response of MMs responds to structural manipulation of the unit cell.

In this paper, we report an experimental study outlining how the nonlinear response of a THz MM can be controlled solely through manipulation of the structural design of the MM inclusions. Our MM design is based on the common broadside-coupled SRR (BCSRR) [29] and is shown schematically in figure 1a. This design is composed of two layered SRR arrays, oriented to maximize electromagnetic interactions between the two arrays. More details on device design are discussed below. We show that by altering the lateral positions of the two component split ring resonators in the unit cell, one can control the field confinement (FC) in the capacitive gaps of the resonators and thus control the coupling of incident THz fields to the nonlinear carrier dynamics of an n-doped GaAs substrate. When the two resonator layers are aligned to overlap as shown in figure 1b, very little nonlinear behavior is seen in the device response at the design frequency of 0.4 THz. When a lateral offset is placed between the upper and lower SRR array (shown in Figure 1c) the FC in the capacitive gap of the lower resonator increases, enabling a nonlinear response from the carrier dynamics of the n-doped epilayer at a relative low incident THz power. The nonlinear carrier dynamics result in a drop in the epilayer conductivity, turning onthe lowest order SRR resonance at 0.4 THz. The resonance is thus highly dependent on incident THz power and on resonance THz transmission is modulated by approximately 3dB as the incident THz power is increased. Below, we present nonlinear THz spectroscopy measurements to characterize device response and numerical simulations to provide insight into the physical mechanisms for device behavior.

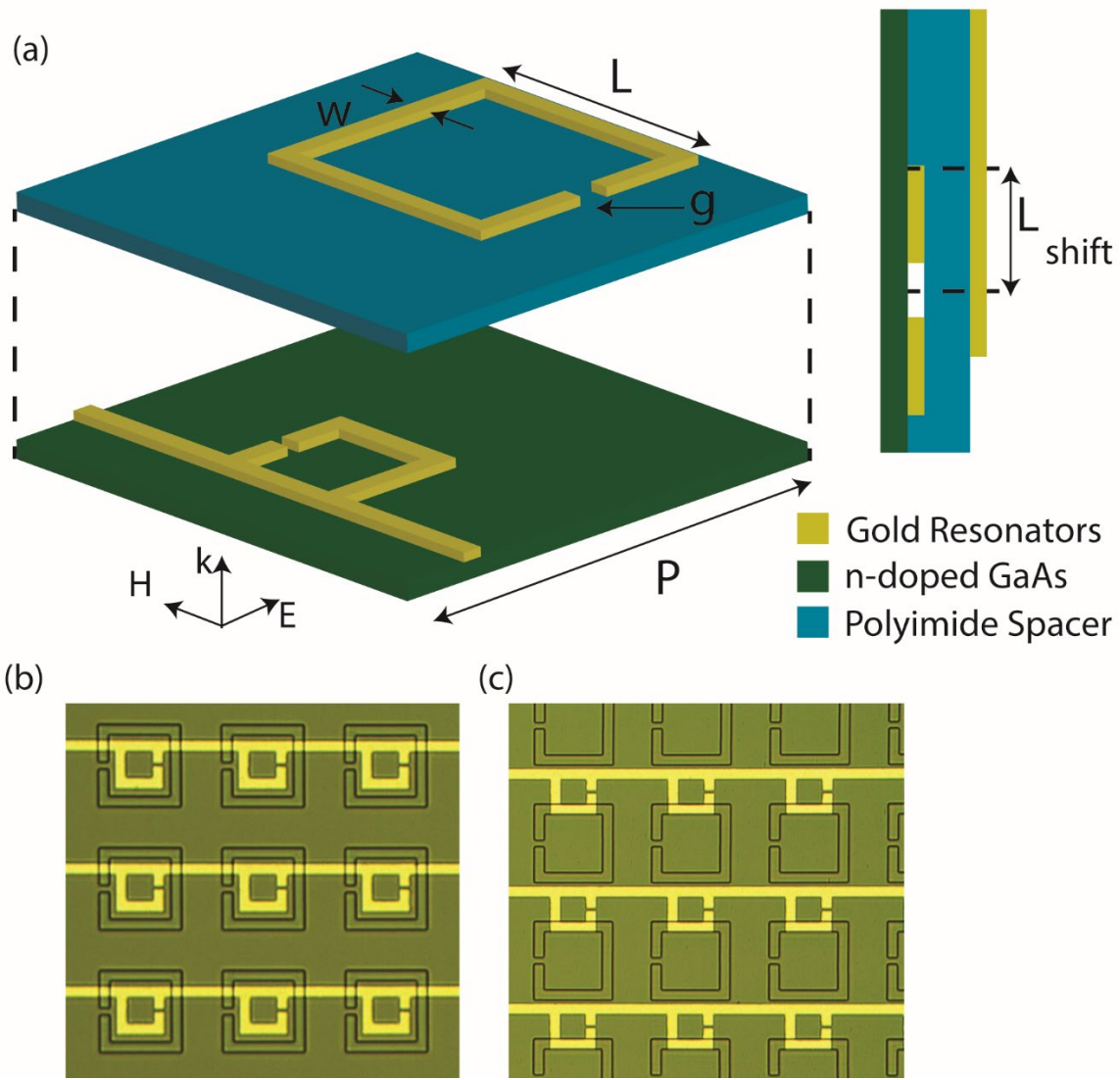

FIG 1: (a) Vertically expanded perspective and side view of metamaterial unit cell showing dimensions and direction of THz excitation. (b) and (c) Mid-fabrication photos of two samples showing a different 0 µm (b) and 48 µm (c) lateral offset between the stacked resonator arrays. The lower layer of SRRs is shown in gold, while the upper layer is shown as outlines.

The MM studied in this work consists of two stacked, planar arrays of gold SRRs as shown in Figure 1. The lower array is fabricated directly onto a GaAs substrate with a 1 µm thick n-doped GaAs epilayer (n

= $2 \times 10^{16}$ cm$^{-3}$) using standard photolithographic techniques. The SRRs of this layer are inter-connected with metallic wires for use in applying an electrical bias (not used in this study). A 2 µm thick polyimide spacer layer is then deposited on the epilayer followed by a second SRR array. The orientation of the second array is rotated by 180° relative to the first array, which maximizes electromagnetic coupling between the two layers.[29]

The dimensions of the component SRRs are chosen to maximize coupling between the two resonator layers and to place the device resonance near the peak of the TPFG THz signal at ~0.4 THz. The key dimensions of each resonator are marked on the upper SRR in figure 1a. These are the side length, L, the capacitive gap width, g, the linewidth, w, and the square array periodicity, P. Below, the dimensions for the lower SRR array are written with a subscript 1, while dimensions for the upper array are written with a subscript 2. In the lower array, $L_1$ = 28 µm, linewidth $w_1$ = 6 µm, and $g_1$ = 2 µm. In the upper array, $L_2$ = 48 µm, $w_2$ = 6 µm, and $g_2$ = 2 µm. The unit cell periodicity is the same for both arrays and is P = 96 µm. Two samples are fabricated each with a varying lateral shift, $L_{shift}$, between the centers of the two SRRs, as shown in the side view of figure 1a. Photographs of the two fabricated samples are shown in figure 1b and 1c. In sample 1 (figure 1b), the two SRR arrays are directly aligned with $L_{shift}$ = 0 µm. In sample 2 (figure 1c), the two arrays are laterally offset by half of the unit cell periodicity ($L_{shift}$ = 48 µm).

To characterize the nonlinear response of the two samples, THz pulses with field strengths between 50 and 400 kV/cm were generated using TPFG in LiNO$_3$ and focused onto the MM at normal incidence with the THz electric field polarized perpendicular to the SRR capacitive gaps, as shown in figure 1a. The transmitted pulses are then measured using electro-optic sampling in ZnTe in a standard THz time-domain spectroscopy (THz-TDS) configuration. The transmission spectra are obtained through Fourier transform and normalized to a THz-TDS reference measurement of a bare GaAs substrate with a 1 µm n-doped epilayer.

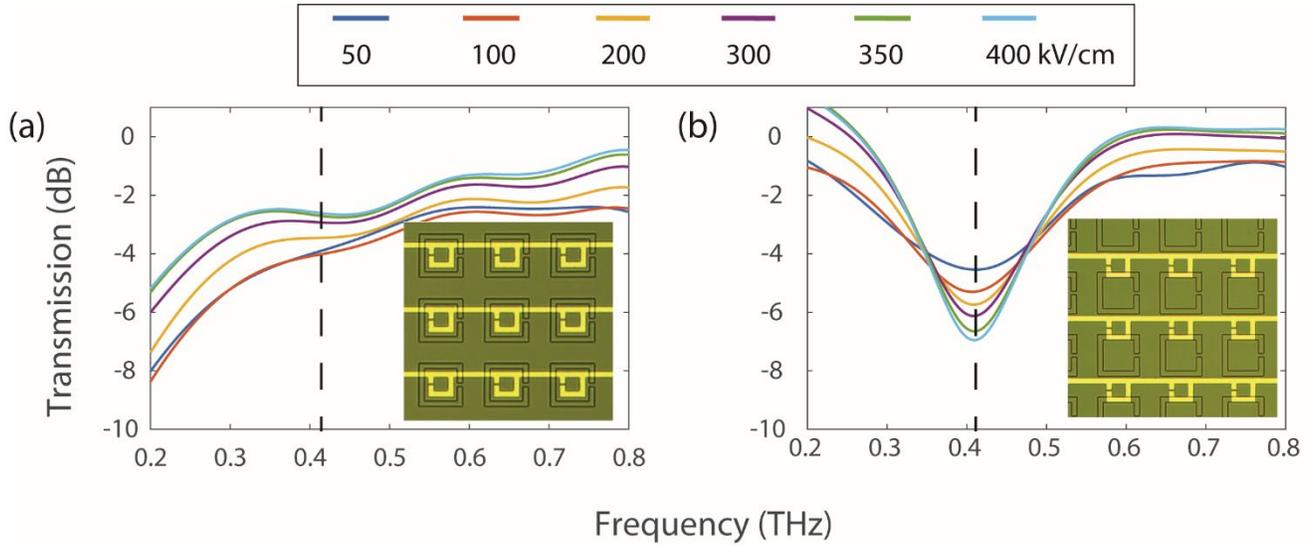

FIG 2: THz-TDS data for samples with (a) $L_{shift}$ = 0 µm and (b) $L_{shift}$ = 48 µm. Insets show photographs of the relative positioning for the two SRR layers for the data shown in the plot. On each plot, the dashed vertical lines mark the position of the resonance of the broadside coupled SRR system.

Experimental transmission spectra as a function of field strength for both samples are shown in figure 2. For the sample with $L_{shift}$ = 0 µm (fig. 2a) only a small nonlinear response can be seen in the data. As the THz field strength is increased from 50 - 400 kV/cm, the overall transmission through the sample slightly increases, but no clear resonance is seen, regardless of incident field strength. Figure 2b shows noticeably different behavior exhibited by the structure in which the SRRs are offset. The MM now exhibits a strong resonance near approximately 0.4 THz with a larger nonlinear modulation. As the THz field strength is increased from 50 - 400 kV/cm, the on-resonance transmission is modulated by approximately 3dB.

Thus, the presence of a strong nonlinear modulation in the MM resonance can be controlled solely by structural positioning of the two MM layers. As another comparison of the stark difference in modulation range, figure 3 plots the transmission at 0.4THz for both samples as a function of incident THz field strength. Not only is the modulation range noticeably increased in the case for $L_{shift}$ = 48 μm, the direction of modulation is also reversed compared to the unshifted structure.

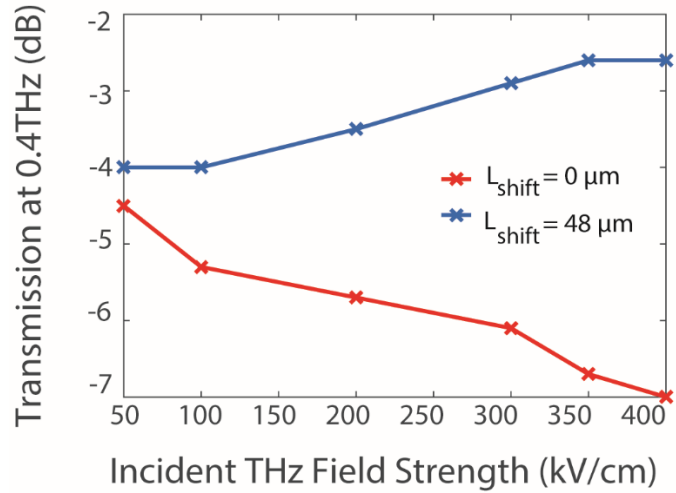

FIG 3: Modulation of the transmission minimum at 0.4THz vs. applied THz field strength for both shifted and unshifted BC-SRR samples. Cross markers on lines signify measured data points.

Using numerical simulations, we investigate the physical origins of the nonlinear response of the BC-SRR MM discussed above. In order to show the mode structure of the MM, we simulate the THz transmission spectra of the BC-SRR structure for $L_{shift}$ = 0 and 48 μm using commercial solvers based on finite difference time domain techniques.[30] The SRR gold patterning is modeled as a lossy metal, while the GaAs substrate is modeled with a 1 μm lossy semiconductor epilayer on a loss-free semi-insulating GaAs substrate. These simulated spectra are shown together in Figure 4a. Due to the limited frequency resolution in the tilted pulse front THz-TDS system, the resonances in simulation are narrower than in the experimental results.

The blue curve in figure 4a shows the spectra for the sample with $L_{shift} = 0$ μm. The incident THz electric field polarized perpendicular to the capacitive gap of the SRRs excites two modes. The resonance at 0.73THz (resonance A) corresponds to the electrically active coupled mode of the BC-SRR. [31] The frequency position and oscillator strength of mode A has been shown in previous work to be highly dependent on the electromagnetic coupling between the two component SRRs. The frequency position of mode A can be approximately described using an LC oscillator model:

$$f_A \sim \frac{1}{2\pi}\left(\frac{1}{\sqrt{LC}}\right) \qquad (1)$$

where L is the BC-SRR total inductance, and C the BC-SRR total capacitance. [32, 33].

The red curve in figure 4a shows the transmission spectrum for the sample with $L_{shift} = 48$ μm. The lateral offset of the two SRRs alters the mutual capacitance and inductance of the structure, red-shifting resonance A from 0.73 THz to 0.47 THz. The mode at 0.66 THz (resonance B) is a surface lattice mode, common in periodic MM structures [34]. The frequency position of mode B is largely independent of the mutual interactions between the two SRRs, as expected for a surface lattice resonance.

Figures 4b and 4c show simulations of the nonlinear response of the $L_{shift} = 0$ μm and $L_{shift} = 48$ μm samples, respectively. Here, the GaAs response is modeled using Drude model with a carrier mobility that can vary between 100 cm²/Vs and 1000 cm²/Vs. The nonlinear response arises from terahertz field induced intervalley scattering of charge carriers in the 1 μm n-GaAs epilayer. The resulting decrease in mobility reduces the conductivity of the epilayer, resulting in a terahertz field dependent increase in resonance depth for resonances A and B for both $L_{shift} = 0$ μm and $L_{shift} = 48$ μm.

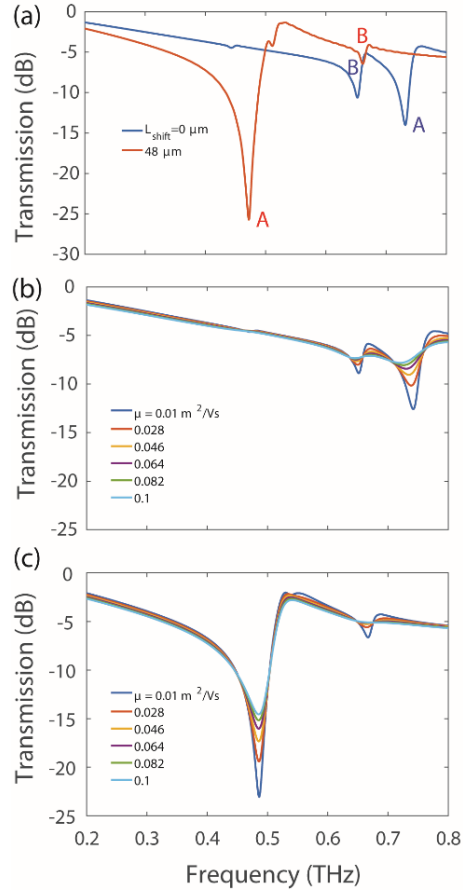

FIG 4: (a) Simulated transmission spectra for the $L_{shift} = 0$ µm and 48 µm samples assuming no loss in the GaAs substrate. Letters A and B denote resonances discussed in the main text. (b) Simulated nonlinear response for $L_{shift} = 0$ µm (c) Simulated nonlinear response for $L_{shift} = 48$ µm.

In experiment, the overall loss in the n-doped epilayer broadens all resonances. The resolution limit of the THz-TDS system also artificially broadens the observed resonances. For $L_{shift} = 0$ µm, neither resonance is seen in experiment (fig. 2a) for any value of incident THz field strength, due to the above-mentioned broadening effects. However, the net decrease in carrier mobility is still seen in the data as a broadband increase in transmission.

For $L_{shift} = 48$ µm (fig 2b), resonance A is clearly visible in the experimental results since the overall oscillator strength, and thus resonance depth, of resonance A is now much greater. Here, the decrease in carrier mobility decreases the on-resonance loss of the BC-SRR structure, leading to a stronger resonance and explaining the difference in modulation direction discussed above in figure 3.

The lateral positioning of the two SRR layers allows for tuning of the oscillator strength by controlling the local FC within the MM unit cell. For mode A, the FC within the capacitive gap region is directly proportional to the oscillator strength of mode A. In addition to the LC resonance model, mode A can be thought of as a resonance of a folded half-wave dipole antenna, where the length, d, of the dipole is the circumference of the SRR. As such, the current distribution in the a BC-SRR resonator can be approximately modeled as a cosine distribution:

$$j = j_o \cos(\frac{\pi s}{d})e^{-i\omega t} \qquad (2)$$

where $j_o$ is the peak current density on resonance, s is the position along the SRR circumference, and $\omega$ is the frequency of incident THz excitation. The current in the upper SRR, oscillates out of phase with the current in the lower SRR. This results in a charge buildup along the gaps of both SRRs, 90º out of phase with the resonant current. Specifically, the charge is:

$$Q = Q_o \sin(\frac{\pi s}{L})e^{-i\omega t} \qquad (3)$$

where $Q_o$ is the peak charge across the SRR capacitive gap. With no lateral offset, the out of phase electric fields from the two resonators superimpose and distructively interfere, leading to a low net electric field strength in the unit cell of the MM. As the two resonators are laterally offset, the resonant electric fields of the two SRRs no longer superimpose spatially, leading to less destructive interference and a higher net electric field strength inside the MM unit cell. The end result is an increase in the strength of the local electric fields in the lower SRR gap region. This results in a stronger resonance for the sample with $L_{shift}$ = 48 μm and a larger nonlinear modulation of the resonance as the incident THz field strength is increased.

We can confirm this explanation of tunable local FC by performing simulations of the local electric field distributions within the MM unit cell. Figure 5 shows the time domain electric field strength maximum in the plane of the lower SRR for the $L_{shift}$ = 0 μm (fig. 5a) and $L_{shift}$ = 48 μm (fig. 5b) samples. The local electric fields are higher by close to a factor of 4 in the $L_{shift}$ = 48 μm sample. As the local electric field amplification increases, especially in the vicinity of the SRR capacitive gap, so will the overall oscillator strength and resonance depth of the BC-SRR resonance.

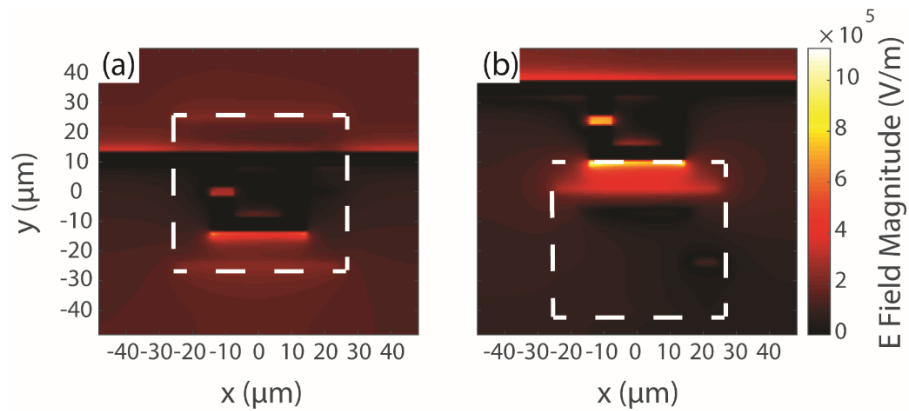

FIG 5: Local electric field distribution in lower SRR during THz excitation for (a)Lshift = 0 μm (b)Lshift =48 μm. White dashed square outlines position of upper SRR layer.

Figure 6a and 6b show simulations of the increase in resonance depth, and thus oscillator strength, of resonance A, for varying values of $L_{shift}$. As $L_{shift}$ is increased from 0 to 48 μm, the depth of resonance A increases by 80%, corresponding to an 80% increase to oscillator strength. Consequentially, resonance A and its nonlinear behavior is visible in the experimental spectra for $L_{shift}$ = 48 μm, but not for $L_{shift}$ = 0 μm.

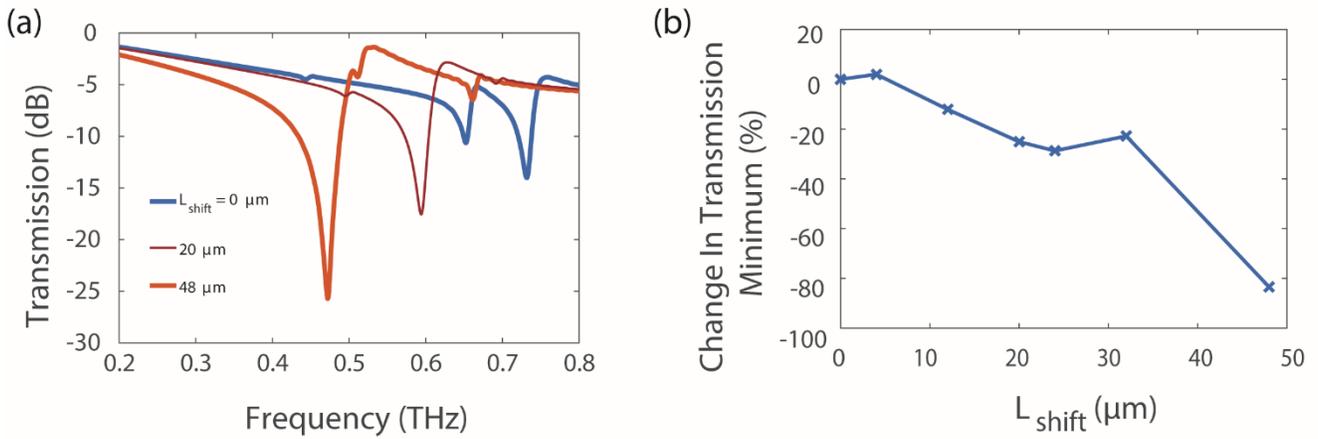

FIG 6: (a) Simulated transmission spectra for BC-SRR samples with varying values of $L_{shift}$. (b) Change in the on-resonance transmission minimum vs. $L_{shift}$. Crosses mark measured data points.

In conclusion, we presented a proof-of-principle study showing how a nonlinear metamaterial response can be tuned in magnitude via the relative lateral positioning of the stacked resonator arrays inside a broadside coupled split ring resonator metamaterial. The metamaterial was patterned on an n-doped GaAs substrate. We investigated the behavior of the metamaterial experimentally via terahertz time domain spectroscopy and use numerical simulations to provide physical insight into the device response. In samples where the two resonator arrays were aligned, the device showed only a small nonlinear response in experiment. In samples where the component split ring resonator arrays were laterally shifted by 48 μm, a resonance appeared at ~ 0.41 THz. Due to the charge carrier dynamics of the n-doped GaAs and on-resonance field confinement in the SRR capacitive gaps, the on-resonance THz transmission was strongly nonlinear, decreasing by approximately 3 dB as the incident terahertz power increases from 50 to 400 kV/cm. This result is, to the best of our knowledge, the first example illustrating the control of the nonlinear response of a THz MM device solely through the structural positioning of the component inclusions.

**Acknowledgements**

This work was supported by the U.S. Department of Energy, Office of Basic Energy Sciences, Division of Materials Sciences and Engineering and performed, in part, at the Center for Integrated Nanotechnologies, an Office of Science User Facility operated for the U.S. Department of Energy (DOE) Office of Science. Sandia National Laboratories is a multi-mission laboratory managed and operated by National Technology and Engineering Solutions of Sandia, LLC, a wholly owned subsidiary of Honeywell International, Inc., for the U.S. Department of Energy's National Nuclear Security Administration under contract DE-NA0003525. Work at UCSD was supported by the DARPA DRINQS program (Grant No. D18AC00014). G.R.K. thanks the Washington College Faculty Enhancement Program for travel and equipment funding throughout the duration of this project.

This paper describes objective technical results and analysis. Any subjective views or opinions that might be expressed in the paper do not necessarily represent the views of the U.S. Department of Energy or the United States Government.

**Data Availability**

The data that support the findings of this study are available from the corresponding author upon reasonable request.